\documentclass[11pt]{article}
\usepackage{moriond,epsfig}

\def\be{\begin{equation}}
\def\ee{\end{equation}}
\def\bea{\begin{eqnarray}}
\def\eea{\end{eqnarray}}

\newcommand{\ups}{$\Upsilon$}
\newcommand{\PbPb}{Pb+Pb}           

\begin{document}
\vspace*{4cm}
\title{Heavy-Ion Physics at the LHC with the Compact Muon Solenoid Detector}

\author{ Christof Roland for the CMS Collaboration}

\address{Massachusetts Institute of Technology, Cambridge, Massachusetts 02139}

\maketitle\abstracts{
Hadron collisions at the LHC offer a unique opportunity to study 
strong interactions.
The exciting data collected by the four RHIC experiments suggest that in heavy-ion collisions at $\sqrt{s_{_{\it NN}}} =$ 200 GeV,
an equilibrated, strongly-coupled partonic system is formed. An extrapolation of the existing data toward LHC energies suggests
that the heavy-ion program at the LHC is in a situation comparable to that of the high-energy program, where new discoveries near
the TeV-scale are expected. Similarly, heavy-ion studies at the LHC are bound to either confirm 
the theoretical picture emerging from RHIC or challenge and extend our present 
understanding of strongly interacting matter at extreme densities. 
The experience at RHIC shows that the ideal detector for future heavy-ion studies should 
provide large acceptance for tracking and calorimetry, high granularity, high resolution and use 
fast detector technologies as well as sophisticated triggering. The CMS detector at the LHC excels in each of these categories. 
}

\section{Introduction}

Heavy-ion collisions at the LHC will explore strongly interacting matter at 
higher densities, higher temperatures and longer lifetimes than ever before. 
The high collision energies at the LHC provide a new set of probes that
are, at best, available with low statistics at presently accessible energies. Examples include 
very high $p_T$ jets and photons, {\rm Z} bosons, 
the $\Upsilon$, {\rm D} and {\rm B} mesons and high-mass dileptons. 
These probes will provide new and quantitative diagnostic tools for the study
of the dense matter produced in heavy-ion collisions. 

\section{The CMS Detector and Heavy-Ion Collisions}
\begin{figure}[t]
\begin{center}
\includegraphics[width=12cm]{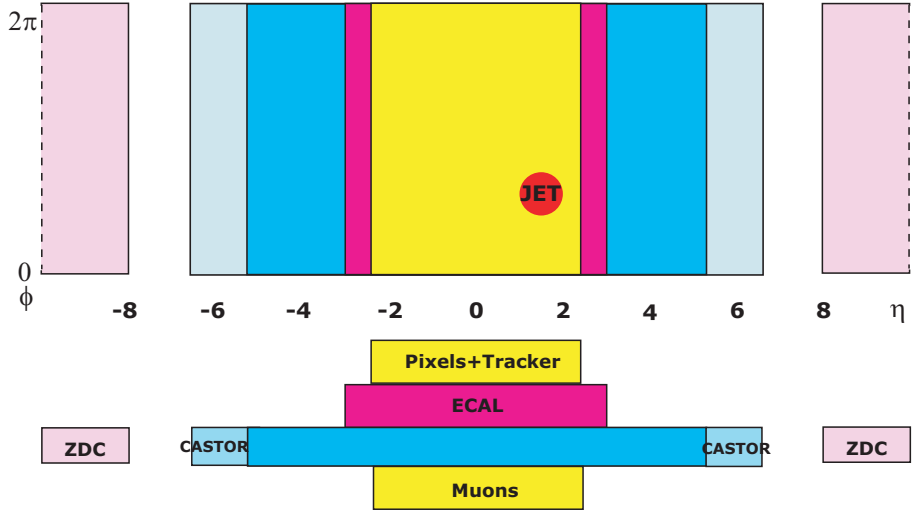}
\caption{ \label{fig-acceptance}
Acceptance of tracking, calorimeters and muon identification in pseudorapidity and azimuth.
The size of a jet with cone radius $R=0.5$ is also depicted as an illustration.}
\end{center}
\end{figure}

The discoveries in the first years of RHIC operation have not only transformed the current 
understanding of nuclear matter at extreme densities, but also greatly shifted the emphasis 
in the observables best suited for extracting the properties of the initial 
high-density QCD system. 
Future studies of heavy-ion collisions and particle physics at very high luminosity accelerators 
require large acceptance, high rate and high resolution detectors. 
The following list illustrates the assets which make the CMS apparatus an 
ideal state-of-the-art heavy-ion detector.

\begin{enumerate}
\item {\bf High Rate:} CMS is designed to deal with {\rm pp} collisions at luminosities of 
up to $10^{34}$ cm$^{-2}$s$^{-1}$, corresponding to {\rm pp} collision rates of $40$~MHz. 
Accordingly, the fast detector technologies chosen for tracking 
(Si-pixels and strips), electromagnetic and hadronic calorimetry and muon 
identification will allow CMS to be read out with a minimum bias trigger at 
the full expected \PbPb\ luminosity. This fast readout will allow
detailed inspection of every event in the high level trigger farm. 

\item {\bf High resolution and granularity:}
The high granularity of the silicon pixel layers, in combination with the 4T magnetic 
field, gives a world best momentum resolution, $\Delta p_T/p_T <1.5\%$ up 
to $p_T \approx 100$~GeV/$c$. At the same time, a track impact parameter resolution at the event vertex of less than 50 $\mu$m ($<20$~$\mu$m at $p_T > 10$ GeV/$c$) is achieved. 
The calorimetry provides 16\% jet energy resolution for 100 GeV
jets with a background charged multiplicity of $dN/dy = 5000$.  
The ECAL spatial resolution in $\eta$ and $\phi$ is 0.028 and 0.032, correspondingly. 
\item {\bf Large acceptance tracking and calorimetry:}
CMS offers high resolution tracking and calorimetry over a uniquely large range 
in pseudo rapidity ($\eta$) and 2$\pi$ in azimuth ($\phi$). The acceptance of the tracking detectors, 
calorimeters and muon chambers can be seen in Fig.~\ref{fig-acceptance}. 
In addition, CMS proposed the CASTOR Calorimeter and the T2 Silicon detector to 
extend the acceptance out to very large rapidity ($\vert \eta \vert$ up to 7).
\item {\bf Particle identification: }
At the LHC, charm and bottom quarks will be copiously produced. 
The large acceptance, high resolution muon system, in combination with the 
tagging of secondary decays by the silicon tracker, will provide the opportunity to 
study the interaction of not only identified hadrons, but also identified quarks with the medium. 
In addition, the physics of meson vs.\ baryon production at large $p_T$ 
can be studied using the results for reconstructed $\pi^0$s, as well as the information provided
by the silicon tracker in combination with the electromagnetic and hadron calorimeters. 
\end{enumerate}
\section{Physics Studies}

The physics program of the Compact Muon Solenoid (CMS) 
encompasses many aspects of heavy ion physics.  The evaluation of simulated data 
indicates that the CMS detector will be well suited for 
{\it (i)} event-by-event charged particle multiplicity and energy flow measurements 
as well as azimuthal asymmetry \cite{CMS2000_60};
{\it (ii)} production of quarkonia and heavy quarks \cite{yellow_report_quarkonia};
{\it (iii)} high $p_T$ particles and jets, including detailed studies of jet fragmentation, jet shapes and jet$+$jet, jet$+\gamma$ and jet$+${\rm Z} correlations \cite{yellow_report_jets}; 
{\it (iv)} energy flow measurements in the very forward region, including neutral and
charged energy fluctuations \cite{CASTOR}; 
{\it (v)} studies of ultraperipheral collisions \cite{CMS2000_60};
{\it (vi)} comparison studies of {\rm pp}, $pA$ and $AA$ collisions.

\begin{figure}[t]
\begin{center}
\begin{minipage}{7.0cm}
\begin{center}
\epsfig{file=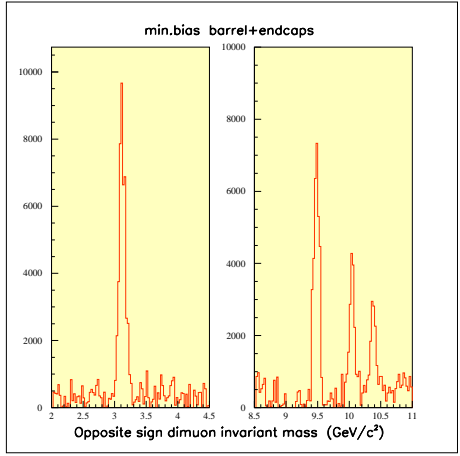,width=6.0cm}
\caption{
Opposite sign dimuon invariant mass.
Left Panel: The $J/\psi$ mass range.
Right panel: Mass range of the \ups~family
}
\label{quarkonia}
\end{center}
\end{minipage}
\end{center}
\end{figure}

\subsubsection{Quarkonia Physics}

The study of the properties and yields of quarkonia 
is an important part of the LHC heavy-ion program.
To date, CMS has focused on the detection of these resonances through their decays to muon pairs.
The muon momentum resolution of the CMS detector translates into
a \ups\ mass resolution of 50 MeV/$c^2$, which provides
a clear separation between the \ups\
family members and a high signal-to-background ratio, as shown in Fig.~\ref{quarkonia}.
The yields are large, allowing studies of
resonance production as functions of $p_T$, rapidity and centrality,
with high statistics in a one month run.

\subsubsection{Hadron yields and jet structure}

\begin{figure}[t]
\hspace{0.3cm}
\begin{minipage}{7.2cm}
\epsfig{file=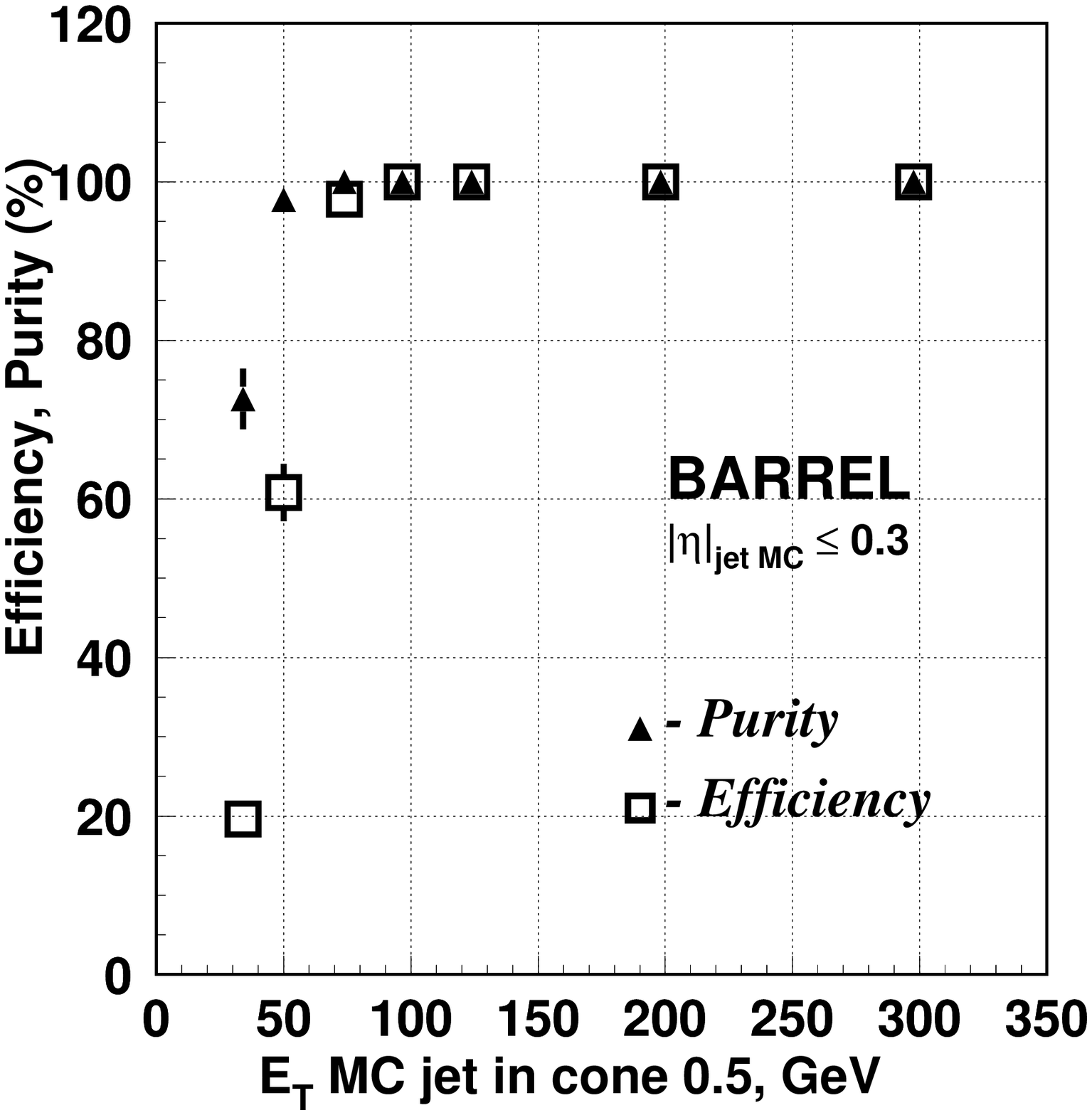,width=6.5cm,height=5.0cm}
\caption{Jet reconstruction efficiency and purity using calorimeters for PYTHIA generated jets embedded in dN/dy = 5000 background events.}
\label{jetreco}
\end{minipage}
\hspace{0.3cm}
\begin{minipage}{7.2cm}
\epsfig{file=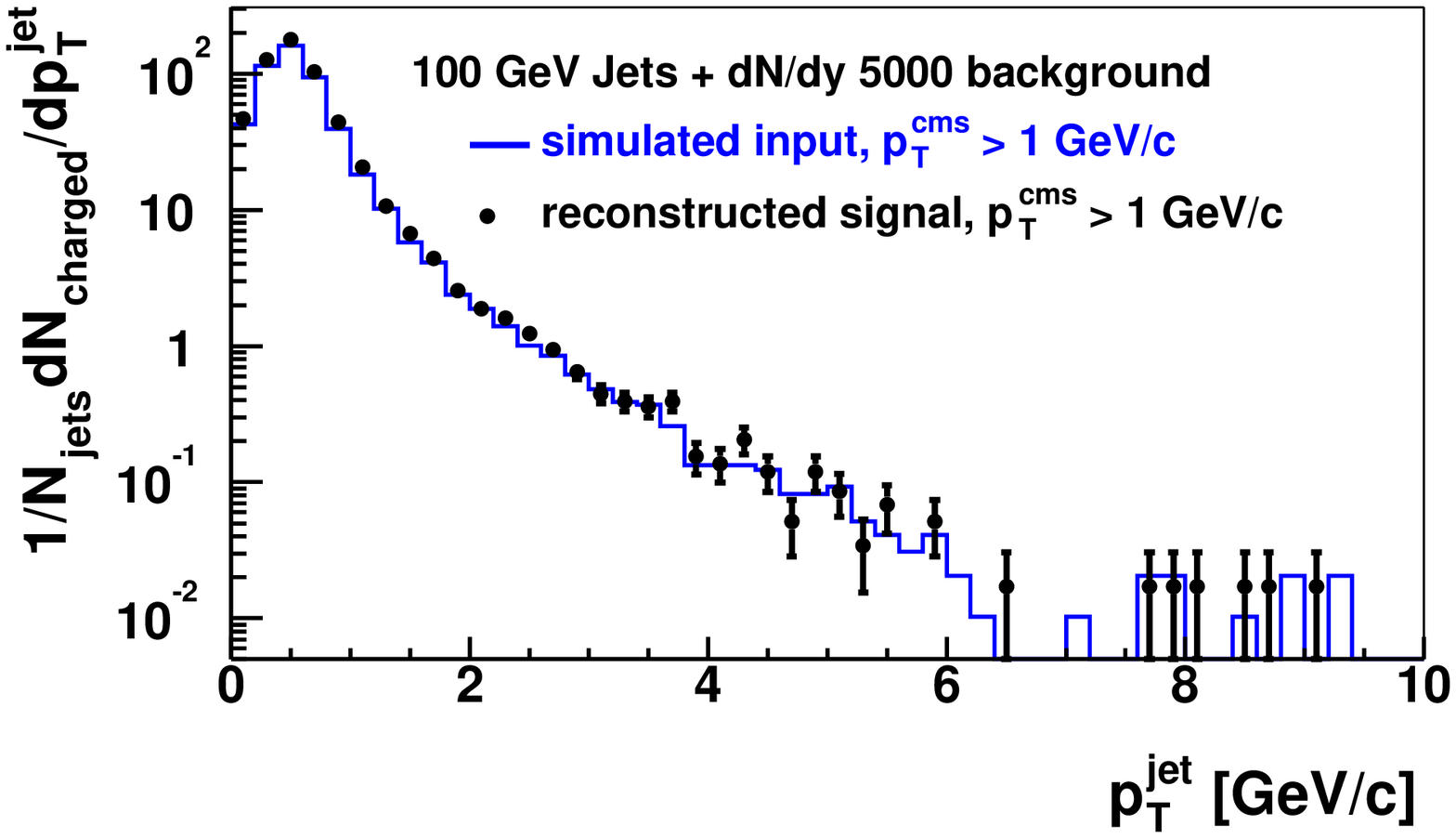,width=6.5cm,height=5.0cm}
\caption{Transverse momentum of charged particles relative to the jet axis, $p_{T}^{{\rm jet}}$, in a 100 GeV jet. A cut on the transverse momentum in the centre-of-mass system, $p_{T}^{{\rm CMS}}$, of 1 GeV/$c$ has been applied.}
\label{jetkt}
\end{minipage}
\end{figure}

Recent results from RHIC \cite{zajc,harris} on the
suppression of the hadron yields above $p_T \geq 3$
GeV/$c$, and the reduction of back-to-back hadron correlations
based on the underlying nucleus-nucleus
collisions \cite{star_jet}, indicate a pronounced energy loss of fast partons. 
The absence of these effects in central {\rm d}Au collisions
suggests that the suppression is an effect of the dense medium created
during the collision. Due to the large increase in the yield of high
$p_T$ hadrons at the LHC, suppression studies can be extended to higher $p_T$
and to fully formed jets.
In addition to the yield suppression, jet
fragmentation and jet shape are expected to be modified by the
presence of the hot medium. 
At the LHC, $10^7$ dijets with $E^{\rm jet}_T> 100$ GeV are produced in 
$|\eta|<$2.6 over a one month Pb+Pb run \cite{yellow_report_jets}. 
The number of dijets is reduced 
by about a factor of two if only the
barrel is considered, $|\eta|<$1.5. 
High energy jets appear as localized energy deposits in the calorimeters.
The jet energy and direction are reconstructed using a iterative cone
type jet-finding
algorithm modified to include background subtraction \cite{CMS2000_60}. The jet-finding efficiency and purity are shown in Fig.~\ref{jetreco}. Even jets 
with energies as low as $50$ GeV
can be reconstructed with good efficiency and low
background using the calorimeters.

The CMS high resolution silicon tracker, together with its pixel detector, allows to 
reconstruct charged particles of  $p_T>1$ GeV/$c$ with good efficiency, low
levels of contamination and precise momentum reconstruction at the highest 
particle densities expected at the LHC. The $p_T$ distribution of particles reconstructed
within an embedded $100$ GeV jet with respect to jet axis is shown in Fig.~\ref{jetkt}. 
The presence of a plasma is expected to
modify this distribution compared to jets produced in {\rm pp} collisions \cite{yellow_report_jets}.

The good performance of the CMS apparatus allows the study of 
jet quenching in dijet production \cite{igor_l}.
Ideal tests of direct jet energy loss arise from processes where a hard parton jet is tagged by an ``unquenched'' ({\it i.e.} not strongly interacting) 
particle such as a {\rm Z} or $\gamma$. Given the known initial parton energy 
measured by the photon, it is possible to perform energy calibrated 
studies of the properties of the quark jets on the opposite side.

\section{Summary} 

The CMS detector is a unique tool to study heavy ion collisions at the LHC.
It probes hot matter through studies of {\rm J}$/\psi$ and $\Upsilon$
production rates.
Its excellent calorimetry and high resolution tracker provides
large coverage and good energy resolution for jet quenching studies. These 
capabilities have been extensively simulated and evaluated.  In addition,
studies of its capabilities for event-by-event charged particle multiplicity,
particle flow, and  jet fragmentation indicate superb detector performance.

\end{document}